\documentclass[prl,twocolumn,showpacs,superscriptaddress]{revtex4}

\usepackage{graphicx}
\begin{document}

\title{MR Imaging of Reynolds Dilatancy in the Bulk of Smooth Granular Flows}

\author{Ken Sakaie} \affiliation{The Cleveland Clinc, 9500
Euclid Ave., Mailcode U-15, Cleveland, OH 44195, USA}

\author{Denis Fenistein} \affiliation{Kamerlingh Onnes Lab,
Universiteit Leiden, Postbus 9504, 2300 RA Leiden, The
Netherlands}

\author{Timothy J. Carroll} \affiliation{Departments of Biomedical
Engineering and Radiology, Northwestern University, 676 N. St.
Clair, Suite 1400, Chicago, Il 60611, USA}

\author{Martin van Hecke} \affiliation{Kamerlingh Onnes Lab,
Universiteit Leiden, Postbus 9504, 2300 RA Leiden, The Netherlands}

\author{Paul Umbanhowar}
\affiliation{Department of Mechanical Engineering, Northwestern
University, Evanston, Illinois 60208, USA}

\date{\today}

\begin{abstract}

Dense granular matter has to expand in order to flow, a phenomenon
known as dilatancy. Here we perform, by means of Magnetic
Resonance Imaging, direct measurements of the evolution of the
local packing density of a slow and smooth granular shear flow
generated in a split-bottomed geometry. The dilatancy is found to
be surprisingly strong. The dilated zone follow the region of
large strain rate and slowly spreads as a function of time. This
suggests that the local packing density is governed by the total
amount of local strain experienced since the start of the
experiment.

\end{abstract}


\maketitle

%Tonight: draft paper , make which list figs. Make webpage for paul
%and mail...

%2) Artificial 2D vel? At least center

%3) densi (time) (exclude surface area). Rate of dens -> guestimate
%for strain rate profile. Do they look ok?

Granular media, such as sand, are conglomerates of dissipative,
athermal particles that interact through repulsive and frictional
contact forces, and that jam into a disordered configuration when
no external energy is supplied \cite{jamming}. When slowly
sheared, granulates flow, but to do so, they must overcome
geometrical (steric) hindrance. The resulting expansion of the
material is referred to as Reynolds dilatancy
\cite{reynolds,caveat}. Precise studies of dilatancy are hindered
by the opaque nature of granular media, and by the complexity of
slow grain flows, which in experimental situations never form a
simple linear shear profile \cite{gdr_leolin,mueggenburg,depken}
--- measurements of average densities therefore cannot probe
dilatancy. Hence, many basic questions are left unanswered. How
much do granular material dilate in the flowing zone? Does the
dilatancy depend on the local strain rate? How far does the
dilated zone extend? Does the packing density evolve over time?

Here we address these issues by imaging the 3D packing density of
smooth and slow granular flows by means of Magnetic Resonance
Imaging (MRI). The flows are generated in a split-bottomed shear
cell, where the grain flow is driven by the rotation of a bottom
disc with respect to a cylindrical container (Fig.~1a). In these
systems, a wide shear zone emanates from the edge of the disc,
making them well suited for studies of smooth and slow granular
flows
\cite{depken,fenistein1,fenistein2,unger,cheng,fenistein3,depken2,unger2,unger3}.
In earlier studies, the ratio of the filling height $H$ to the
radius of the disc $R_s$ was found to govern the qualitative
nature of the flow
\cite{fenistein1,fenistein2,unger,cheng,fenistein3,unger2}: When
$H/R_s$ is less than $0.6$, the central core material rests on and
rotates as a solid with the disc. The shear zone then reach the
free surface, and the three dimensional shape of the region of
large shear resembles the cone of a trumpet
\cite{fenistein2,unger,cheng,fenistein3,unger2}. When $H/R_s
\gtrsim 0.75$, the flowing zone do not reach the surface but forms
a ``dome'' shaped shear zone in the bulk of the material
\cite{unger,cheng,fenistein3,unger2}. For intermediate filling
depths there is a crossover regime \cite{fenistein3}.

By imaging the local packing densities in our shear cell as a
function of time and for a range of filling heights, we explore
the relation between flow field and dilatancy. The simple picture
that emerges is that the amount of dilatancy grows with the total
amount of local strain experienced and saturates when the local
strain becomes of order one. First, we  show that the relative
change in density in the flowing zone is rather strong and
saturates around 10-15 \%. Secondly, we find that the dilated zone
slowly spread throughout the system as time progresses. This
spread is consistent with the idea that the total, local strain
experienced since the start of the experiment governs the amount
of dilatancy --- to show this, we will reconstruct the flow field
in the bulk, combining previously found scaling relations
\cite{fenistein2,unger,depken,depken2} with measurements of the
velocity at the surface. Finally, and in contrast to small filling
heights ($H/R_s<0.6$) where the locations of the dilated zone and
the shear zone coincide, for deep filling heights a relatively
long-lived transient causes the dilated zone to deviate from the
late-time shear zone.
%The density here probes
%behavior which is very difficult to access otherwise.

{\em Experimental setup ---} By detecting the signal of protons in
oil within seeds, MRI has been used to non-invasively probe the
dynamics of granular materials
\cite{cheng,Nakagawa,Ehrichs,mueth}. While various techniques,
based on either spin-tagging or employing mixtures of grains with
distinct oil concentration, have been used to track grain motion,
we are not aware of prior research employing MRI to measure
packing densities $\phi$. Here, food grade poppy seeds sieved to
select diameters of one mm were used.  Multiple horizontal slices
were acquired in an interleaved fashion to reduce signal
saturation effects \cite{mridetails}. Slices were 2 mm thick with
in-plane resolution of 1.56 mm x 1.56 mm. Individual seeds are
thus not resolved, but the MRI signal probes local packing
densities. To compensate for the overall gradients in the signal,
we compare signal intensity before and after flow has occurred,
and report the relative change in density $\Delta \phi \equiv
\phi(t)/\phi(0)-1$. We employ the azimuthal symmetry of the system
to average the data and focus on $\Delta \phi (N,r,z)$, where
$N,r$ and $z$ denote the number of rotations of the disc since the
start of the experiment and the radial and vertical coordinate
respectively.

The experimental geometry placed into the MRI apparatus is a
stationary plexiglass drum of inner radius 95 mm in which the
bottom disk of radius $R_s = 70$ mm is slowly rotated with a rate
$\Omega$ of approx 1/8 $s^{-1}$ (see Fig.~1). At this rate, the
flow velocities are simply proportional to $\Omega$
\cite{depken,fenistein1,fenistein2,unger,cheng,fenistein3,depken2,unger2,unger3,mueth}.
The cell is filled with poppy seeds and tapped to obtain a
well-compacted initial condition of filling depth $H$. A layer of
seeds is bonded to the side walls and bottom to obtain rough
boundaries. To study the temporal evolution of the density, we
have obtained images after $N$ rotations of the bottom disc.
During the MRI procedure, the disc is stopped
--- starting and stopping the flow has no appreciable effect
on these slow, non-inertial flows. To facilitate direct
comparisons to the flow field, we used video imaging to obtain the
non-dimensional surface velocity profile $\omega(r)$, defined as
the ratio of the time averaged azimuthal velocity to the driving
rate $\Omega$ \cite{fenistein1,fenistein2,fenistein3}.  We focus
on late times after transients have died out.

\begin{figure}[tb]
\includegraphics[width=8.5cm]{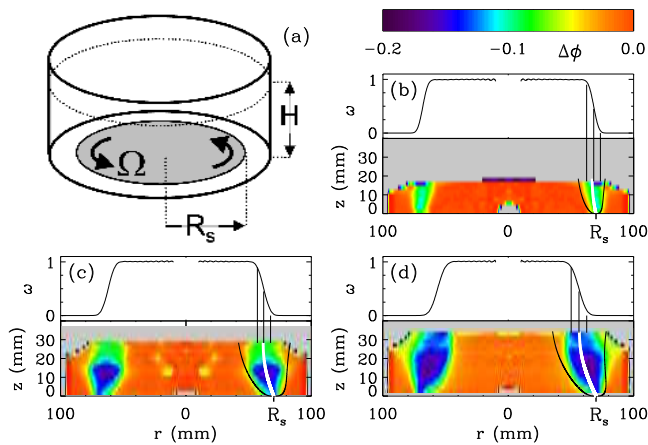}
\caption{(Color) (a) Sketch of the split-bottomed shear cell, in
which a layer of granular material of depth $H$ is driven by the
rotation of a disc of radius $R_s$ with rate $\Omega$. (b-d).
Comparison of the surface flow field $\omega(r)$ and the packing
density in the bulk after $N$ rotations. (b) $H/R_s=0.23$  and
$N\!=\!2.5$; (c) $H/R_s=0.41$ and $N\!=\!4$; (d) $H/R_s=0.51$ and
$N\!=\!4$. The straight, vertical lines indicate where $\omega(r)$
reaches 0.1, 0.5 and 0.9, thus indicating the location of the
shear zone as observed at the free surface. The white curves are a
theoretical prediction for the center of the shear zone based on
Eq.~\ref{rcbulk} \cite{unger}. The black curves show examples of
the edge of the shear zone, based on Eq.~\ref{rcbulk} and
\ref{wbulk}.}\label{fig1}
\end{figure}

{\em Dilated zone and Shear zone for low filling heights ---}
Figure~1 illustrates shear and dilated zones are directly related
for low filling heights ($H/R_s \lesssim 0.55$). First, near the
surface, the location of the dilated zone and the shear zone
coincide. Measurements of the surface velocity profiles
$\omega(r)$ are in agreement with earlier studies
\cite{fenistein1,fenistein2,cheng,fenistein3,depken2} since we
find that
\begin{equation}\label{erf}
\omega(r) = 1/2 + 1/2 ~\mbox{erf}\left[ (r-R_c)/W\right]~,
\end{equation}
where the center location $R_c$ in consistent with the simple
scaling law:
\begin{equation}\label{rcsurf}
1 - R_c/R_s =(H/R_s)^{5/2}~.
\end{equation}
Figure 1b-d shows that the centers of the dilated and shear zones
coincide at the free and bottom surfaces, and that the width of
both zones grows similarly with $H/R_s$.

Second, in the bulk the dilated zone also follow the shear zone.
The center of the shear zone throughout the bulk $R_c$ and the
vertical coordinate $z$ are related to the geometrical parameters
$H$ and $R_s$ by a scaling argument as \cite{unger}:
\begin{equation}\label{rcbulk}
z=H-R_c (1-R_s/R_c(1-(H/R_s)^{2.5}))^{0.4}~.
\end{equation}
Fig.~1b-d shows that this curve falls into the center of the
dilated zone.

Third, data on the width of the shear zone in the bulk, obtained
by either excavating colored particles
\cite{fenistein2,fenistein3} or MRI and numerical studies of the
flow field \cite{cheng,depken2}, indicate that the shear zone
rapidly widen with distance from the bottom disc as \cite{footung}
\begin{equation}\label{wbulk}
W \sim z^{1/3}~.
\end{equation}
We have shown examples of such curves in Fig.~1b-d, and find that
they describe the shape of the dilated zone well
\cite{footung,footw}. Overall, our data clearly suggest that for
$H/R_s \lesssim 0.55$ the shear and dilated zones are directly
related.

{\em Spread of the dilated zone ---} Notwithstanding their close
relation, the shear zone and dilated zone are not identical. To
clarify the relation between dilatancy and flow field further, we
show in Fig.~2a the evolution of the density profiles obtained
along a number of vertical slices as a function of the number of
rotations $N$.

We first focus on the density profiles for fixed $N$. The local
flow and strain rate vary substantially throughout the shear zone
(see below), but in contrast, the dilatancy clearly reaches a
plateau value throughout the shear zone. Packing density and local
strain rate are thus not directly related. We have found that this
maximal dilatancy is similar for all our data sets, and does not
vary much with time and depth, except close to the free surface
\cite{footw}. The amount of dilatancy is substantial, with the
densities in the plateau 10-15\% lower than the non-sheared
regions. To put this number in perspective, disordered packings of
spheres span a range in packing densities from 55\% (Random Loose
Packing) to 64 \% (Random Close Packing), with typical densities
of poured grains around 60 \% \cite{Onada,Weitz}. Even though
poppy seeds are not perfectly spherical (see Fig.~2b), so that the
precise values of these numbers may be somewhat different, the
observed dilatancy is substantial.

\begin{figure}[tb]
\includegraphics[width=8.5cm]{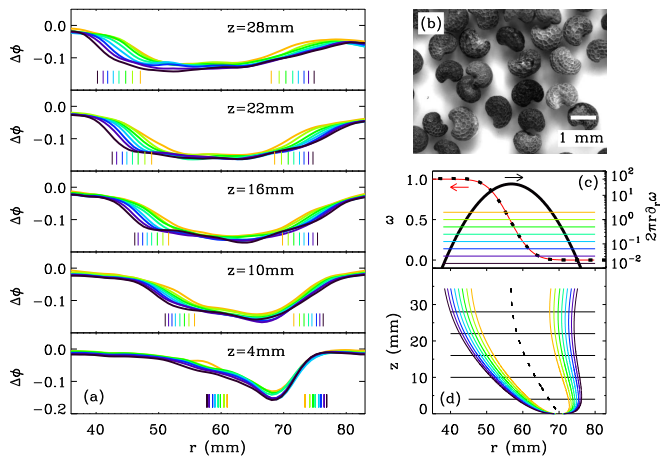}
\caption{(Color) (a) Horizontal profiles of the density changes
for the data set with $H/R_s\!=\!0.51$ (shown in Fig.~1d), as
functions of $N$ and for five values of $z$ as indicated. $N \!=\!
1/2,1,2,4,\dots 64$, with $N\!=\!1/2$ shown in orange, and
$N\!=\!64$ in black. The vertical lines indicate estimates of
where, after $N$ turns, the local strain rate equals one. (b)
Snapshot of the poppy seeds used in this experiment. (c) Surface
velocity $\omega(r)$ (dashed), a fit to the error function form of
Eq.~(\ref{erf}) (red), and the local strain after one rotation
$2\pi r \partial_r \omega$ (black, log scale). The horizontal
lines indicate where the local strain $\gamma$ reaches one after
$N$ rotations
--- colors as in panel (a). (d) Illustration of the locations
where $\gamma$ after $N$ rotations reaches one, constructed from
the strain at the surface and the scaling laws Eq.~(\ref{rcbulk})
(dashed line) and Eq.~(\ref{wbulk}). }\label{fig2}
\end{figure}

The most striking feature  shown in Fig.~2a is that the dilated
zone slowly spreads with the number of rotations $N$
\cite{spreadnote}. As we will now show, the picture that emerges
is that the total amount of strain, not the strain rate, governs
the amount of dilatancy --- entirely consistent with a
quasi-static picture of slow grain flows.

This is illustrated in Fig.~2. Away from the container bottom,
vertical gradients are small, and we approximate the total local
strain in the bulk after $N$ turns by
\begin{equation}
\gamma = 2 \pi N r \partial_r \omega ~.
\end{equation}
To obtain $\partial_r \omega$, we start from the surface velocity
profile (black dashed curve in Fig.~2c), to which we fit the error
function form of Eq.~(\ref{erf}) (red curve in Fig.~2c) before
taking the derivative to obtain the local strain after one
rotation ($2\pi r
\partial_r \omega$) (black curve in Fig.~2c, log scale).
The horizontal lines in Fig.~2c indicate where the local strain at
the surface after $N$ rotations reaches one.

Figure~2d illustrates how we determine the local strain field in
the bulk from the surface measurements. We start with the
observation that the functional form of $\omega(r)$ in the bulk is
also an error function \cite{cheng,fenistein3}. The center
location of this shear zone is given by Eq.~\ref{rcbulk} (dashed
curve in Fig.~2d). Finally, in the bulk the width of the shear
zone scales as $z^{1/3}$
\cite{fenistein2,cheng,fenistein3,depken2}, so using the measured
width of the shear zone at the free surface we can calculate the
complete strain rate field. From this we can then deduce, for any
number of turns $N$, where in the bulk the local strain reaches
one (colored curves in Fig.~2d).

In Fig.~2a, the density profiles for $N=1/2,1,2,4,\dots,64$ are
compared to the location in the bulk where after $N$ turns $\gamma
\!=\!1$. We see that, at least qualitatively, the location where
the strain reaches one coincides with the edge of the dilated
zone. Near the container bottom, the coincidence is less good, but
there vertical gradients, deviations from Eq.~\ref{wbulk} and
boundary effects may be substantial. Nevertheless, our data
provides clear evidence that the spreading of the dilated zone is
simply caused by the spreading  of the position where the local
strain becomes of order one with time. This suggests that there is
no steady state and that the dilated zone  continually expands
since the strain rate is nowhere strictly zero \cite{spreadnote}.

In prior work focussing on the velocity field for similar filling
heights, no long time evolution was observed
\cite{fenistein1,fenistein2,cheng,fenistein3,depken2}. This does
not necessarily  disagree with the slow dilatancy. To measure
velocities accurately, one requires grains to move substantially
in order to average out fluctuations. However, the system is
already fully dilated after a local strain of order $1$, so that
transients in the flow field caused by changes in the density
escape detection
--- essentially, dilatancy happens in the far tails of the
velocity profiles \cite{unger3}.

\begin{figure}[t]
\includegraphics[width=8.5cm]{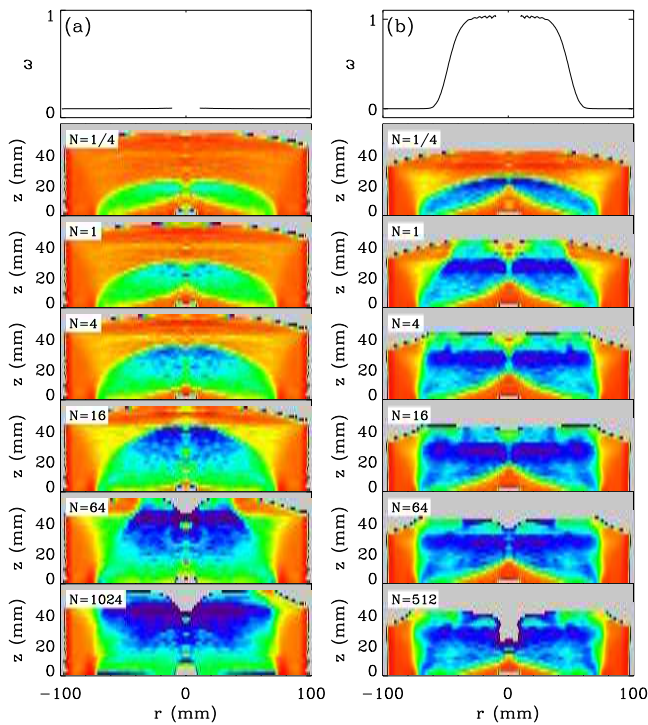}
 \caption{(Color)
 Surface velocity profiles and dilatancy maps for (a) $H/R_s=0.77$
and (b) $H/R_s=0.61$. See text.}\label{fig3}
\end{figure}

{\em Dilated zone for large filling heights --- } In Fig.~3a we
show the surface velocities and density evolution for $H/R_s \!=\!
0.77$. For such deep filling depths, the shear zone emanating from
the edge of the disk meet in the middle of the cell and form a
dome-like structure, and the surface remains essentially
stationary \cite{unger,cheng,fenistein3,unger2}. Simple scaling
laws for the flow field like Eqs.~(\ref{erf}-\ref{wbulk}) are not
known, but the qualitative features of the dilated zone at early
times also show the dome-like shape
\cite{unger,cheng,fenistein3,unger2}. For later times the dilated
zone spreads throughout large parts of the system, leading to an
appreciable elevation and corrugation of the free surface. Slow
convection, occurring for large filling heights, leads to the slow
growth of a dip in the center of the free surface
\cite{cheng,fenistein3,cheng_private_com}.

In Fig.~3b we show the surface velocities and density evolution
for $H/R_s \!=\! 0.61$. This corresponds to the intermediate
regime, where the velocity profile is weakly asymmetric
\cite{fenistein3}, but there is no precession. We recover this
asymmetry of $\omega(r)$, and also note that the center of the
surface remains stationary during a short transient. The bulk
density evolution reveals that the 3D flow profile is a complex
mix between the ``trumpet'' shape observed for small filling
heights and the ``dome'' shape observed for large filling heights
\cite{cheng,fenistein3,cheng_private_com}. For early times, all
the dilatancy is concentrated in a dome like structure, consistent
with the steady surface for early times. For late times, the shear
zone evolves to the trumpet shape leading to $\omega \approx 1$
near the center of the surface. We suggest that the symmetry
breaking of the surface flow profiles and the dome-like dilated
zone might be related \cite{fenistein3}. Also here a dip develops
near the free surface (the almost vertical shape of this dip for
$N=512$ is an imaging artefact).

{\em Discussion --- } Our data uncovers how granular media dilate
under steady shear \cite{oscifoot}: the amount of dilatancy grows
with the accumulated strain, and saturates when this strain
becomes of order one. This fully dilated state might very well
coincide with what in the engineering literature is known as the
``critical state'', referring to a final state of grain flow where
rheological properties stay constant \cite{unger3}. How general
this correspondence is remains to be seen, in particular for shear
bands near a boundary or free surface \cite{gdr_leolin,mueth}.

DF acknowledges support from FOM, and MvH acknowledges support
from NWO/VIDI and discussions with X. Cheng.

~


\begin{thebibliography}{99}
\bibitem{caveat} Note that very loosely packed materials may
actually compact upon shearing.
\bibitem{footung} In \cite{unger3}, the authors find a somewhat
different scaling law for the width of the shear zone. With the
present MRI resolution, results based on this scaling are not
significantly better than those based on Eq.~\ref{wbulk}.
\bibitem{footw}
Surprisingly, near the free surface the dilated zone appears to
shrink in width, in qualitative contrast to what has been observed
for the shear zone. This appears to be a general feature of the
density maps we obtained, and while we cannot exclude imaging
artifacts near the free surface, we believe that this is caused by
a difference in the initial packing density near the surface ---
presumable, deep inside the layer where the grains sustain a
larger weight, the material is slightly more compacted than near
the surface, so that the latter show a smaller relative change in
density.
\bibitem{spreadnote} We have found evidence for continued dilated zone
spreading after hundreds of rotations of the bottom disc.
\end{thebibliography}
\end{document}